\documentclass{iopart}

\usepackage{iopams}
\begin{document}


\newtheorem{defn}{Definition}

\nosections

\newtheorem{teo}[defn]{Theorem}
\newtheorem{eje}{Example}
\newtheorem{lem}[defn]{Lemma}
\newtheorem{rem}[defn]{Remark}
\newtheorem{cor}[defn]{Corollary}
\newtheorem{pro}[defn]{Proposition}
\newtheorem{property}{Property}

\makeatother \font\ddpp=msbm10  at 11 truept
\def\R{\hbox{\ddpp R}}
\def\C{\hbox{\ddpp C}}
\def\L{\hbox{\ddpp L}}
\def\S{\hbox{\ddpp S}}
\def\Z{\hbox{\ddpp Z}}

\newcommand{\D}{{\cal D}}
\newcommand{\M}{{\cal M}}
\newcommand{\Mo}{{\cal M}_0}
\newcommand{\la}{\Lambda}
\newcommand{\dimo}{{\bf Proof: }}
\newcommand{\inte}{\int_{0}^{1}}
\newcommand{\gam}{\gamma}
\newcommand{\eps}{\epsilon}
\newcommand{\<}{\langle}
\renewcommand{\>}{\rangle}
\newcommand{\Om}{\Omega^1}
\renewcommand{\(}{\left(}
\renewcommand{\)}{\right)}
\renewcommand{\[}{\left[}
\renewcommand{\]}{\right]}
\newcommand{\om}{\omega}
\newcommand{\me}{\frac{1}{2}}
\newcommand{\Mt}{\widetilde{\M}}
\newcommand{\cat}{{\mathop{\rm cat}\nolimits}}

\newcommand{\cvd}{\ \rule{0.5em}{0.5em}}

\newcommand{\bm}[1]{\mbox{\boldmath $#1$}}

\newcommand{\be}{\begin{equation}}
\newcommand{\ee}{\end{equation}}
\newcommand{\noi}{\noindent}
\newcommand{\ben}{\begin{enumerate}}
\newcommand{\een}{\end{enumerate}}
\newcommand{\bit}{\begin{itemize}}
\newcommand{\eit}{\end{itemize}}
\newcommand{\edoc}{\end{document}}

\article[Uniqueness of static decompositions]{NOTE}{A note on the uniqueness of global static decompositions}

\author{Miguel S\'anchez$^{1}$ and
Jos\'e M M Senovilla$^{2}$}
\address{$^1$ Departamento de Geometr\'{\i}a y Topolog\'{\i}a,
Facultad de Ciencias, Universidad de Granada,
Avenida Fuentenueva s/n, 18071 Granada, Spain}
\address{$^2$ Departamento de F\'\i sica Te\'orica e Historia de la Ciencia,
Facultad de Ciencia y Tecnolog\'{\i}a, Universidad del Pa\'{\i}s Vasco,
Apartado 644, 48080 Bilbao, Spain}
\eads{\mailto{sanchezm@ugr.es}, \mailto{josemm.senovilla@ehu.es}}

\begin{abstract}
We discuss when static
Killing vector fields are standard, that is, leading to a {\em
global} orthogonal splitting of the spacetime. We prove that such
an orthogonal splitting is {\em unique} whenever the natural space
is compact. This is carried out by proving that many
notable spacelike submanifolds must be contained in an orthogonal slice.
Possible obstructions to the global splitting are also considered.
\end{abstract}

\maketitle



The main goal of this short Note is to analyze the uniqueness of the typical orthogonal splitting, adapted to a static Killing vector field, of static spacetimes. Along the way, we will also
clarify some technical points concerning static spacetimes, such as (i) when
the existence of a static Killing vector field yields a {\em global} decomposition of the spacetime ---which are  then called ``standard static''; and (ii)
possible obstructions to the existence of
standard static spacetimes.

Our main result is Theorem \ref{tuniq}, stating that any standard
static spacetime with a compact space ---defined as the set of
trajectories of the static Killing vector field--- has a unique
global decomposition of the form of expression (\ref{ems}) below.
For non-unique splittings with non-compact spaces see \cite{Tod}.

Let $(M,g)$ be a spacetime,
i.e., a connected smooth $n(\geq 2)$-dimensional manifold $M$
endowed with a time-orientable $C^2$ Lorentzian metric $g$. A time-orientation is fixed and
will be implicitly used.  Our notation and conventions are standard,
as in the books \cite{BEE,O,SW}, see also
\cite{Sa-nonlin} for a review.

A Killing vector field $\vec\xi$ will be called:
\ben
\item {\em (Strictly) stationary} if $\vec\xi$ is timelike (and then, by reversing the sign if necessary, can be assumed to be future-directed).

\item {\em Static} if $\vec\xi$ is stationary and irrotational, usually also called hypersurface-orthogonal \cite{Exact}, i.e. its orthogonal distribution is involutive or, equivalently, $\bm{\xi} \wedge d\bm{\xi} =0$, where $\bm{\xi}$ is the one-form associated to $\vec \xi$ via the metric by ``lowering" the index: $\bm{\xi}\equiv g(\vec\xi , \cdot)$.

\item {\em Standard static} if $\vec\xi$ is static and the full spacetime $(M,g)$ is isometric
to a product manifold $\R\times S$ endowed with the metric
\be
\label{ems}
 g= -V^2 dt^2 + g_S ,
\ee
where $\vec\xi =\partial_t$, the function $V$ is such that $\vec\xi(V) =0$, and $g_S$ is a truly Riemannian metric on $S$, therefore independent of $t$.

Note that $V^2 = -g(\vec \xi,\vec \xi)$ and that $(S,g_S)$ is
isometric to any integral manifold of the orthogonal distribution
of $\vec\xi$. Thus the splitting (\ref{ems}) is univocally
determined, up to isometries, by the standard static Killing
vector field $\vec\xi$. The Riemannian space $(S,g_S)$ can be also
identified with (a) any of the hypersurfaces $t=$const., which are
orthogonal to $\vec\xi$, endowed with its natural first
fundamental form and (b) the quotient space $(M/\!\!\sim , g_S)$
where $\sim$ is the equivalence relation defined by the integral
curves of $\vec\xi$ so that two points in $M$ are equivalent if
they belong to the same integral curve. (Recall also that there
are general results ensuring the completeness of Killing vector
fields \cite{Chrus} and that the quotient by the flow of a
complete stationary Killing vector field is a Hausdorff manifold
\cite{Harris}).
\een
Accordingly, the spacetime will be called {\em stationary, static
{\em or} standard static}. Observe that
the splitting (\ref{ems}) 
may be
non-unique in an essential way: apart from the trivial $a\vec \xi$ with constant $a>0$, other independent standard static Killing vector fields may exist (for instance, $\vec\xi =\partial_t$ and $\vec\xi'=2\partial_t+\partial_x$ in Minkowski spacetime $\L^4$ in usual Cartesian coordinates). We will later prove, however, that this is impossible if $S$ is compact.

Locally, any static spacetime looks like a standard one, with the (chosen) static Killing
vector field identifiable to $\partial_t$.
In this (non-necessarily standard) static case, the
integral hypersurfaces of the foliation orthogonal to $\vec\xi$ do make sense, but they do not have to be achronal, nor necessarily homeomorphic to $M/\!\!\sim$, as shown in the following example.
\begin{eje}
Take for instance the 2-dimensional cylinder $\R\times S^1$ with $g=-dt^2+d\theta^2$ and the (non-standard) static Killing vector field $\vec\eta=2\partial_t+\partial_\theta$. Its orthogonal hypersurfaces are topologically $\R$, while the quotient space is topologically $S^1$.
\label{ex1}
\end{eje}

Some remarkable and worth-mentioning properties of stationary and static spacetimes are the following (see \cite[Sect.3]{Sa-nonlin} for further information):

\begin{property} {\em The Killing vector fields on $(M,g)$ form a finite-dimensional Lie algebra, and
the stationary ones constitute a convex subset: if $\vec\xi_1, \vec\xi_2$
are stationary, $\lambda \xi_1 + (1-\lambda) \xi_2$ is stationary for all
$\lambda\in [0,1]$. Nevertheless, static vector fields do not form a convex
subset; in fact, the sum $\vec\xi_1 + \vec\xi_2$ of static $\vec\xi_1 ,
\vec\xi_2$ may be non-static. For example, consider  the region $|t|<x$ of
$\L^4$, and take $\vec\xi_1= 2\partial_t +
\partial_y$, $\vec\xi_2= x\partial_t + t\partial_x$.}
\end{property}

\begin{property} {\em As shown in \cite[Th. 2.1(1)]{Sa-dga}
{\em any complete static vector field in a
simply connected spacetime is standard static}. Furthermore,
in dimension $n=2$, any stationary vector field is
static, for any 1-dimensional distribution is involutive.}
\label{property2}
\end{property}

\begin{property} {\em Concerning geodesic completeness and related issues, we have the following results.
(a) The orthogonal hypersurfaces of any static Killing vector field are totally geodesic, as their second fundamental form vanishes.
(b) In the particular standard static case (\ref{ems}), if $(M,g)$ is
geodesically complete then so is $(S,g_S)$. The converse
does not hold, as illustrated by the simple example $(\R^2,g=-e^x dt^2 +dx^2)$.
(c) Non-standard static vector fields may be incomplete, even in the cases where there is another standard (ergo complete) static Killing vector field (take the strip $-1<x<1$ in $\L^2$).
(d) Nevertheless, if $(M,g)$ is complete all its Killing vector fields
are complete \cite[Prop. 9.30]{O}. Actually, if the spacetime is only null or timelike geodesically complete, then any stationary Killing vector field is complete \cite[Lemma 1]{GH}.
(e) The completeness of $(M,g)$ is ensured by general results in
\cite{RS} (see \cite[Th. 2.1]{Sa-nonlin}). In particular, if $S$
is compact the standard static spacetime, as well as all its
Killing vector fields, are complete.}
\label{property3}
\end{property}

Before proceeding to prove our main results,
let us make some additional comments on the stationary case.
In the expression (\ref{ems}) there are no crossed terms between
the $\R$ and $S$ parts. If these terms appear and are
independent of the variable $t$, the spacetime as well as
the corresponding Killing vector field $\vec\xi\equiv
\partial_t$ are called {\em standard stationary}. But, as a
difference with the static case, {\em a standard stationary vector
field does not imply a unique splitting of the spacetime as a
standard stationary one}. For example, if the spacetime is globally hyperbolic, any
complete stationary vector field $\vec\xi$ is standard stationary. In
fact, a standard stationary splitting is obtained by taking any
spacelike Cauchy hypersurface $S$ (which must exist by
\cite{BS-cmp1}) and moving it by means of the flow of $\vec\xi$ . But
there are many, even non-isometric, choices of $S$ (for example, $S$
can contain any prescribed compact acausal spacelike submanifold,
see \cite{BS-lmp}), each one yielding a different splitting.
On the other hand, a standard stationary spacetime (even globally
hyperbolic) which admits a complete static vector field may just be
non-standard static, see \cite[Sect. 3]{Sa-nonlin}.


We start by proving two results on the non-existence of certain
submanifolds in a standard static spacetime.
Recall that a (possibly degenerate)
imbedded\footnote{
This can be relaxed to immersed submanifolds, but the details are cumbersome and add very little to the reasoning.} submanifold $N$ of a
spacetime $(M,g)$ is called {\em totally geodesic} if any geodesic
$\gamma$ of the spacetime with initial velocity tangent to $N$
remains in $N$. In this case, $N$ is complete if so are all such geodesics. A non-denegerate $N$ is totally geodesic if and only if all geodesics of the first fundamental form in $N$ are also geodesics of $(M,g)$.

\begin{pro}
\label{telemental} Let $(M=\R \times S,g)$ be a standard static
spacetime (\ref{ems}) with upper bounded $V$. Any totally geodesic
and geodesically complete submanifold $N$ is either
\ben
\item fully contained in a slice $t=$constant, or
\item such that
$t$ attains all possible real values.
\een
\end{pro}

{\em Proof.} If there is a tangent vector $\vec v\in TN$ not
tangent to any $t=$constant slice, take the corresponding geodesic
$\gamma$ with initial velocity $\vec v$. As $\partial_t$ is a
Killing vector, its scalar product with the tangent vector field
to $\gamma$ is a constant, say $c$. Obviously, $c\neq 0$ because
$\vec v$ is not orthogonal to $\partial_t$. Therefore,
$$
\frac{d(t\circ \gamma)}{ds}
=  -\frac{c}{V^2}
$$
so that our assumption on $V$ yields
$$
\left|\frac{d(t\circ \gamma)}{ds}\right| \geq a >0
$$
for some positive constant $a$. As $\gamma$ is complete, $t$ runs on the whole $\R$ and the
result follows because $\gamma$ remains in $N$. \cvd

The most relevant case arises when $N$ is
spacelike. In that case, the strong assumption that $N$ has vanishing shape tensor (or
second fundamental form vector, see \cite{O}) $\vec{\bm{K}}$ can be much weakened to just a very mild condition on its mean curvature vector $\vec H\equiv$ tr$\vec{\bm{K}}$.
Simultaneously, however, one must strengthen the hypothesis on completeness by requiring
compactness. The precise statement is

\begin{pro}
\label{th0}
Let $(M=\R \times S,g)$ be a standard static spacetime
(\ref{ems}). Then, any  spacelike compact submanifold $N$ whose mean curvature vector is orthogonal to the standard static Killing vector field $\vec\xi =\partial_t$ is fully contained in a slice $t=$ constant.

This holds in particular if  $N$ is critical, i.e. $\vec H=\vec 0$.
\end{pro}
{\em Proof.}
 Let $\phi : N \hookrightarrow M$ be the imbedding and
denote its pull-back by $\phi^*$. As $\vec\xi$ is a Killing vector
field it follows (see e.g. \cite[formula (2)]{M-S}) that
$$
g(\vec\xi,\vec H)=-\mbox{div} (\phi^*\bm{\xi})
$$
where div is the divergence operator associated to the first
fundamental form in $N$. From our assumption we derive
div$(\phi^*\bm{\xi})=0$ and, taking into account that {\em
staticity} implies $\bm{\xi}=-V^2dt$, we get \be
\mbox{div}\left[(V\circ\phi)^2 d(\phi^*t)\right]=0. \label{pde}
\ee
Expanding $\mbox{div}\left[(\phi^*t)\, (V\circ\phi)^2 d(\phi^*t)\right]$, integrating then on $N$, using Gauss' Theorem and (\ref{pde}) we arrive at
$$
\int_N (V\circ\phi)^2 \left|d(\phi^*t)\right|^2=0
$$
where the norm is computed with respect to the first fundamental form in $N$.
This immediately implies that $d(\phi^*t)$ must vanish
identically (i.e., the only solutions to the {\em elliptic}
partial differential equation  (\ref{pde}) for $\phi^*t$ are
constant). \cvd

\begin{rem} {\em Several remarks are in order:
\begin{itemize}
\item Technically, the previous result and its proof are essentially contained in \cite[Prop.
2]{M-S}, where the standard static character of $\vec\xi$ was implicitly assumed.
\item Observe, more importantly, that this result can be generalized to non-compact $N$ whenever $V$ satisfies some appropriate decaying properties.
\item Proposition \ref{th0} can be regarded as a generalization to the Lorentzian
case of the following well-known result:
{\em there are no compact minimal surfaces in $\R^3$}. If such a surface existed,
and $\phi: N \hookrightarrow \R^3$ denotes its inmersion, the
components would satisfy $(\Delta \phi^1, \Delta \phi^2, \Delta \phi^3)= \vec H$,
where $\Delta$ is the Laplacian in $N$ with the induced
metric. So, the absurd conclusion that $\phi^i \equiv $ constant for
$i=1,2,3$ would follow.
\end{itemize}
}
\end{rem}
In the case of hypersurfaces (codimension 1), Proposition
\ref{th0} is meaningful only in the maximal case $\vec H = \vec
0$. However, this is no restriction at all, because in this case
there are stronger results. To start with, in any {\em stationary}
spacetime, closed spacelike submanifolds with a future- (or past-)
pointing mean curvature vector field are forbidden unless $\vec H
=\vec 0$ everywhere \cite[Theorem 1]{M-S}. Consider then the case
of a closed hypersurface $N$, and let $\vec n$ be its (say)
future-directed normal vector field, so that  $\vec H = \theta
\vec n$. It follows from the above that $\theta$ cannot be
non-positive, nor non-negative, unless $\theta =0$ on $N$. But
then Proposition \ref{th0} implies that $N$ is actually fully
contained in a slice $t=$ constant, that is to say, $N$ is one of
these slices; hence, $N$ is not only critical, but also totally
geodesic:

\begin{cor} In a standard static spacetime (\ref{ems}), the only possible closed hypersurfaces with $\theta\geq 0$ or $\theta\leq 0$ (including the constant mean curvature case) are the canonical $t$=constant slices. Therefore, they are  totally geodesic.
\end{cor}

Our main uniqueness result is a consequence of Propositions \ref{telemental} or \ref{th0}.
\begin{teo}
\label{tuniq}
Let $(M=\R \times S,g)$ be a standard static spacetime
as in (\ref{ems}), with standard static Killing vector field $\vec \xi =\partial_t$.
If $S$ is compact then any other standard static vector field is of the form
$a\vec\xi$ where $a>0$ is a positive constant.

Thus, the splitting (\ref{ems}) is unique, up to the trivial re-scaling
$\tilde t = t/a, \tilde V =a V$.
\end{teo}
{\em Proof.} Suppose that there exists another standard static
vector field $\vec\xi'$. If $\vec\xi'$ were not parallel to
$\vec\xi$ at some point $p$, its orthogonal hypersurface $S'$
passing through $p$ would not be contained in the slice $t= t|_p$
of the standard splitting associated to $\vec\xi$. But this would
contradict any of propositions \ref{telemental} or \ref{th0}
because $S'$ must be compact (in fact, homeomorphic to
$S$) and totally geodesic.  Thus, $\vec\xi'$ and $\vec\xi$ are
collinear everywhere. As they are both Killing vector fields, they
must be proportional $\vec\xi'=a\vec\xi$ with constant $a$. \cvd

\begin{rem} \label{r3}  {\em Observe that Theorem \ref{tuniq} forbids the
existence of an independent {\em standard} static Killing vector field, but
not the existence of other independent static ones (which, moreover,
must be necessarily complete due to property \ref{property3}(e)). Example \ref{ex1} serves to illustrate this point, for
$\partial_t$ is standard static while $\vec\eta$ is static but not standard.}
\end{rem}

Despite the previous Remark \ref{r3}, additional results on non-existence
of non-standard static vector fields can be easily derived, such as the next corollary.
\begin{cor}
\label{cuniq}
Let $(M=\R \times S,g)$ be a standard static spacetime
as in (\ref{ems}), with standard static Killing vector field $\vec \xi =\partial_t$.
If $S$ is compact and simply connected then any other static Killing
vector field $\vec\xi'$ takes the form $a\vec\xi$ with positive constant $a$.
\end{cor}
{\em Proof.} By the uniqueness Theorem \ref{tuniq}, and the
equivalence between complete static and standard static Killing vector
fields in the simply connected case (property \ref{property2}), one only
has to ensure that $\xi'$ is complete. But this is a consequence
of the completeness of $g$ (property \ref{property3}(e)). \cvd

\begin{rem} {\em Assume that  simple connectedness is not
imposed in Corollary \ref{cuniq}, and $\xi'$ is a (necessarily
non-standard) static vector field, with orthogonal integral
manifold $S'$ (possibly non-compact, as in Example \ref{ex1}) and
flow $\Phi'$. The restriction of $\Phi'$ to $\R \times S'$ is a
covering map on $M$ (use \cite[Cor. 7.29]{O}). Moreover, assume
that $M$ satisfies (vacuum or more general) Einstein field equations. If
$\xi'$ is used to pose Killing initial data on $S$, then $M$ will
be recovered, but if $\xi'$ is used on $S'$ (say, as in \cite{BT})
then $\R\times S'$, endowed with the pullback metric, will be
obtained. }
\end{rem}

One obvious problem is left open: what about  the case with
non-compact slices? It is known that there exist spacetimes with
two different splittings of type (\ref{ems}), such as $\L^n$ or
other non-trivial cases \cite{Tod}. However, all these cases are
very special, having a specific Petrov type, and a high degree of
symmetry. We expect that our techniques can be extended to
determine the full list of spacetimes violating the uniqueness of
the global splitting (\ref{ems}).

\ack{MS is partially supported by the Spanish MEC-FEDER Grant MTM2007-60731
and the Junta de Andalucia Regional Grant P06-FQM-01951. JMMS
is supported by grants
FIS2004-01626 of the Spanish CICyT and GIU06/37 of the University
of the Basque Country.}

\end{document}